\global\def\draftcontrol{0}

   \def\versionno{ susysin}

\catcode`\@=11

\expandafter\ifx\csname draftcontrol\endcsname\relax\global\def\draftcontrol{0}
\fi

{\count255=\time\divide\count255 by 60
\xdef\hourmin{\number\count255}
\multiply\count255 by-60\advance\count255 by\time
\xdef\hourmin{\hourmin:\ifnum\count255<10 0\fi\the\count255}}
\def\draftdate{\number\month/\number\day/\number\year\ \ \ \hourmin }

\newcommand\makepapertitle{\par
  \begingroup
    \renewcommand\thefootnote{\@fnsymbol\c@footnote}%
    \def\@makefnmark{\rlap{\@textsuperscript{\normalfont\@thefnmark}}}%
    \long\def\@makefntext##1{\parindent 1em\noindent
            \hb@xt@1.8em{%
                \hss\@textsuperscript{\normalfont\@thefnmark}}##1}%
     \newpage
     \global\@topnum\z@   
     \@makepapertitle
     \thispagestyle{empty}\@thanks
  \endgroup
  \setcounter{footnote}{0}%
  \global\let\thanks\relax
  \global\let\makepapertitle\relax
  \global\let\@makepapertitle\relax
  \global\let\@thanks\@empty
  \global\let\@author\@empty
  \global\let\@date\@empty
  \global\let\@title\@empty
  \global\let\title\relax
  \global\let\author\relax
  \global\let\date\relax
  \global\let\and\relax
  \def\version{\let\version\@version\@gobble}
}
\def\@makepapertitle{%
  \newpage
   \ifnum\draftcontrol=1 {}
   \version\versionno
   \vskip 3em%
   \else
   \hfill\hbox to 3cm {\parbox{4cm}{\@pubnum}\hss}%
   \vskip 3em%
   \fi
   \begin{center}%
   \let \footnote \thanks
     {\LARGE {\@title}}%
     \vskip 1.5em%
     {\normalsize
       \lineskip .5em%
       \begin{tabular}[t]{c}%
         \@author
       \end{tabular}\par}%
     \vskip 1.5em%
     {\@bstract}%
     \end{center}%
     \vskip 1.5em
     \@date%
   \par
}

\gdef\@pubnum{}
\def\pubnum#1{%
  \gdef\@pubnum{#1}}

\gdef\@bstract{}
\def\Abstract#1{%
  \gdef\@bstract{%
   \parbox{\textwidth-0pc}{%
   \centerline{\bf Abstract}\penalty1000%
\kern.2cm%
\noindent
\renewcommand\baselinestretch{1.0}%
{#1}}}
}

\def\ps@paper{\let\@mkboth\@gobbletwo%
     \ifnum\draftcontrol=1
    \def\@oddfoot{\hbox to \textwidth{\tiny \versionno \hfil\tiny\draftdate}%
    \hskip -\textwidth \hbox to \textwidth{\hfil\rm\thepage\hfil}}%
     \else\def\@oddfoot{\hbox to \textwidth{\hfil\rm\thepage\hfil}}
     \fi
     \let\@evenfoot\@oddfoot
}

\def\body{\clearpage
          \pagestyle{paper}
    }

\def\@version#1{\ifnum\draftcontrol=1
\typeout{}\typeout{#1}\typeout{}
\vskip3mm\centerline{\hbox{\fbox{\normalsize{\tt DRAFT -- #1 -- }
                   {\draftdate}}}}\vskip3mm
\fi}
\let\version\@version
\long\def\eqlabel#1{\ifnum\draftcontrol=1
                    \tag@false  
                    \tag*{(\theequation) \hbox to -0.2cm{\hspace{0cm}\small{#1}\hss}}
                    \refstepcounter{equation}
                    \edef\@currentlabel{\theequation}
                    \ltx@label{#1}          
                    \else
                    \label{#1}
                    \fi
                    }
\let\st@bibitem\@bibitem
\let\st@lbibitem\@lbibitem
\ifnum\draftcontrol=1
  \def\@bibitem#1{%
    \st@bibitem{#1}\a@@label{#1}\ignorespaces}
  \def\@lbibitem[#1]#2{%
    \st@lbibitem[#1]{#2}\a@@label{#2}\ignorespaces}
  \def\a@@label#1{%
    \gdef\a@lab{\smash{\normalfont\small#1}}
    \ifvmode
      \if@inlabel
        \global\setbox\@labels\hbox{%
          \llap{\a@lab\let\a@lab\relax
                \kern\@totalleftmargin\kern\marginparsep}%
          \box\@labels}%
      \fi
    \fi}
\fi

\documentclass[12pt,letterpaper]{article}

\usepackage{amsmath,amssymb,array,calc,epsfig,rotating,bm}
\usepackage[sort]{cite}
\usepackage{graphicx}
\usepackage{psfrag,verbatim}

\ifnum\draftcontrol=1
\fi

\renewcommand\baselinestretch{1.25}
\renewcommand\section{\@startsection {section}{1}{\z@}%
                                   {-3.5ex \@plus -1ex \@minus -.2ex}%
                                   {2.3ex \@plus.2ex}%
                                   {\normalfont\large\bfseries}}
\renewcommand\subsection{\@startsection{subsection}{2}{\z@}%
                                   {-3.25ex\@plus -1ex \@minus -.2ex}%
                                   {1.5ex \@plus .2ex}%
                                   {\normalfont\normalsize\bfseries}}
\renewcommand\subsubsection{\@startsection{subsubsection}{3}{\z@}%
                                   {-3.25ex\@plus -1ex \@minus -.2ex}%
                                   {1.5ex \@plus .2ex}%
                                   {\normalfont\normalsize\it}}
\renewcommand\paragraph{\@startsection{paragraph}{4}{\z@}%
                                   {-3.25ex\@plus -1ex \@minus -.2ex}%
                                   {1.5ex \@plus .2ex}%
                                   {\normalfont\normalsize\bf}}

\numberwithin{equation}{section}

\def\revise#1       {\raisebox{-0em}{\rule{3pt}{1em}}%
                     \marginpar{\raisebox{.5em}{\vrule width3pt\
                     \vrule width0pt height 0pt depth0.5em
                     \hbox to 0cm{\hspace{0cm}{%
                     \parbox[t]{4em}{\raggedright\footnotesize{#1}}}\hss}}}}

\newcommand\nxt[1]  {\\\fnxt#1}
\newcommand{\ie}{{\it i.e.,}\ }

\def\cala         {{\cal A}}

\def\calc         {{\cal C}}

\def\cale         {{\cal E}}

\def\calh         {{\cal H}}

\def\calm         {{\cal M}}

\def\calo         {{\cal O}}

\def\calr         {{\cal R}}

\def\del          {\partial}

\def\Re           {{\rm Re\hskip0.1em}}
\def\Im           {{\rm Im\hskip0.1em}}

\def\sqr#1#2{{\vcenter{\vbox{\hrule height.#2pt
 \hbox{\vrule width.#2pt height#1pt \kern#1pt
 \vrule width.#2pt}\hrule height.#2pt}}}}

\def\a{\alpha}

\def\w{\omega}

\def\dd{\delta}

\def\aa1{\phi}
\def\cc1{\psi}

\def\k{\kappa}

\def\l{\lambda}

\def\k{\kappa}

\def\t{\tau}

\def\hw{\hat{\omega}}

\catcode`\@=12

\begin{document}


\title{\bf Ringing in de Sitter spacetime}

\date{July 4, 2017}

\author{
Alex Buchel\\[0.4cm]
\it $ $Department of Applied Mathematics\\
\it $ $Department of Physics and Astronomy\\ 
\it University of Western Ontario\\
\it London, Ontario N6A 5B7, Canada\\
\it $ $Perimeter Institute for Theoretical Physics\\
\it Waterloo, Ontario N2J 2W9, Canada
}

\Abstract{Hydrodynamics is a universal effective theory 
describing relaxation of quantum field theories towards equilibrium.
Massive QFTs in de Sitter spacetime are never at equilibrium.  We use
holographic gauge theory/gravity correspondence to describe relaxation
of a QFT to its Bunch-Davies vacuum --- an attractor of its late-time
dynamics.  Specifically, we compute the analogue of the quasinormal
modes describing the relaxation of a holographic toy model QFT in de
Sitter.
}

\makepapertitle

\body

\version\versionno
\tableofcontents

\section{Introduction}\label{intro}

Isolated strongly interacting systems typically\footnote{There are some exceptions to 
this lore: condensed matter systems with many-body localization \cite{mbl}; holographic models 
with phase-space restricted dynamics \cite{Balasubramanian:2014cja}. } reach a thermal equilibrium state at late times of 
its dynamical evolution. An approach towards equilibrium is governed by hydrodynamics --- a universal effective theory 
organized as derivative expansion of the local velocity gradients to the temperature of the final equilibrium state.   
One example is the relativistic hydrodynamics of conformal gauge theories developed in \cite{Baier:2007ix,Bhattacharyya:2008jc}.
As an effective description, gradient expansion of the gauge theory hydrodynamics 
has zero radius of convergence 
due to the existence of the non-hydrodynamic modes in equilibrium plasma \cite{Heller:2013fn,Buchel:2016cbj}. 
Whenever gauge theory allows for a dual holographic description \cite{m1,Aharony:1999ti} 
in terms of classical supergravity, its thermal equilibrium state  
is represented by a black hole/black brane in the gravitational dual \cite{Witten:1998zw}. 
Furthermore, linearized hydrodynamic and non-hydrodynamic excitations about the equilibrium state are 
mapped to the quasinormal modes (QNMs) of the corresponding 
dual black hole \cite{Berti:2009kk}. QNMs encode the information about the 
relaxation of the near-equilibrium state of a gauge theory plasma    
\cite{Buchel:2015saa,Fuini:2015hba,Janik:2015waa,Buchel:2015ofa}.    

Implicit in the above overview was an assumption that QFT dynamics occurs in Minkowski spacetime. 
Using holographic correspondence\footnote{For early work on gauge theories in de Sitter 
within holographic framework see \cite{Buchel:2002wf,Buchel:2002kj,Buchel:2003qm,Buchel:2004qg,Buchel:2006em}.}, 
it was argued in  \cite{Buchel:2017pto,Buchel:2017qwd} 
that massive gauge theories in de Sitter spacetime are not in equilibrium at late times:
while Bunch-Davies (BD) vacuum is the late-time attractor of a dynamical evolution of a QFT state, 
the co-moving entropy production rate  is nonzero. 
In this paper we make the first step addressing the question:  
\begin{center}
{\it What is the effective theory of the relaxation towards Bunch-Davies 
vacuum of a massive QFT? }
\end{center}

We restrict our attention to a simple holographic toy model of a $2+1$-dimensional massive $QFT_3$ 
with the effective dual gravitational action\footnote{We set the radius $L$ of an asymptotic $AdS_4$ geometry 
to unity.}:
\begin{equation}
\begin{split}
S_4=\frac{1}{2\kappa^2}\int_{\calm_4} dx^4\sqrt{-\gamma}\left[R+6-\frac 12 \left(\nabla\phi\right)^2+\phi^2\right]\,.
\end{split}
\eqlabel{s4}
\end{equation}
The four dimensional gravitational constant $\k$ is related to the ultraviolet (UV) conformal fixed point
$CFT_3$  central charge $c$  as 
\begin{equation}
c=\frac{192}{\k^2}\,.
\eqlabel{c}
\end{equation}
$\phi$ is  a  gravitational  bulk scalar with 
\begin{equation}
L^2 m^2_\phi=-2 \;,
\eqlabel{mphi}
\end{equation}
which is dual to a dimension $\Delta_\phi=2$ operator $\calo_\phi$ of the boundary theory. 
$QFT_3$ is a relevant deformation of the UV  $CFT_3$ with 
\begin{equation}
\calh_{CFT}\ \to\ \calh_{QFT}= \calh_{CFT}+\Lambda\ \calo_\phi \;,
\eqlabel{defformation}
\end{equation}
with $\Lambda$ being the deformation mass scale. We study $QFT_3$ dynamics 
 in de Sitter spacetime with a Hubble constant $H$; thus 
the metric on $\calm_4$ boundary, $ds_{\del\calm_4}^2$, is taken as 
\begin{equation}
ds_{\del\calm_4}^2=-dt^2 +e^{2 H t}\ \left(dx_1^2+dx_2^2\right)\,. 
\eqlabel{delms}
\end{equation}
Following 
\cite{Buchel:2017pto}, in the next section we describe gravitational dynamical setup encoding  de Sitter evolution of 
spatially homogeneous and isotropic states of the boundary field theory.  
We study the late-time attractor of the evolution 
in section \ref{bdv}. In section \ref{ring} we compute the spectrum of linearized 
fluctuations of the boundary theory around its BD vacuum. 
In section \ref{nl} we use fully nonlinear characteristic formulation of asymptotically $AdS$ dynamics
\cite{Chesler:2013lia}
and establish that generic homogeneous and isotropic states of the boundary theory indeed 
``ring-down'' to BD vacuum with frequencies computed in section \ref{ring}. 
 We conclude in section \ref{end}.

\section{Holographic gravitational dynamics}

A generic state of the boundary field theory with a gravitational dual \eqref{s4}, homogeneous and isotropic in the spatial
boundary coordinates $\boldsymbol{x}=\{x_1,x_2\}$, leads to a bulk gravitational metric ansatz
\begin{equation}
ds_4^2=2 dt\ (dr -A dt) +\Sigma^2\ d\boldsymbol{x}^2\,,
\eqlabel{EFmetric}
\end{equation}
with the warp factors $A,\Sigma$ as well as the bulk scalar $\phi$
depending only on $\{t,r\}$. From
the effective action \eqref{s4} we obtain the following equations of
motion:
\begin{equation}
\begin{split}
&0=d_+'\Sigma+d_+\Sigma\ \left(\ln\Sigma\right)'-\frac 32 \Sigma-\frac 14\Sigma \phi^2 \;,\\
&0=d_+'\phi+d_+\phi\ \left(\ln\Sigma\right)'+\frac{d_+\Sigma}{\Sigma}\ \phi'
+\phi\;,\\
&0=A''-2\frac{d_+\Sigma}{\Sigma^2}\ \Sigma'+\frac 12 d_+\phi\ \phi'\;,
\end{split}
\eqlabel{ev1}
\end{equation}
as well as the Hamiltonian constraint equation:
\begin{equation}
0=\Sigma''+\frac 14\Sigma  (\phi')^2\,,
\eqlabel{ham}
\end{equation}
and the momentum constraint equation:
\begin{equation}
\begin{split}
&0=d_+^2\Sigma-2 A d_+'\Sigma-\frac{d_+\Sigma}{\Sigma^2}\
\left(A\Sigma^2\right)'
  +\frac 14\Sigma  \left((d_+\phi)^2 
 +2A\left(6+\phi^2 \right) \right)\,.
\end{split}
\eqlabel{mom}
\end{equation}
In \eqref{ev1}-\eqref{mom} 
we denoted $'= \frac{\del}{\del r}$, $\dot\ =\frac{\del}{\del t}$, 
and $d_+= \frac{\del}{\del t}+A \frac{\del }{\del r}$. 
The near-boundary $r\to\infty$ asymptotic behaviour
of the metric
functions and the scalar encode the mass parameter $\Lambda$ and the boundary
metric scale factor $a(t)\equiv e^{Ht}$:
\begin{equation}
\begin{split}
&\Sigma=a\biggl({r}+\l+\calo(r^{-1})\biggr)\,,\qquad A=\frac{r^2}{2}+\left(\l-\frac{\dot a }{a }\right)r+\calo(r^0)\,,
\qquad \phi=\frac{\Lambda}{r}+\calo(r^{-2})\,.
\end{split}
\eqlabel{bcdata}
\end{equation}
$\l=\l(t)$ in \eqref{bcdata} is the residual radial coordinate diffeomorphism parameter \cite{Chesler:2013lia}.
An initial state of the boundary field theory is specified providing the scalar
profile $\phi(0,r)$ and solving the
constraint \eqref{ham}, subject to the boundary
conditions \eqref{bcdata}. Equations \eqref{ev1} can then be used to evolve
the state.

The subleading terms in the boundary expansion of the
metric functions and the scalar encode the evolution of the  energy
density $\cale(t)$, the pressure $P(t)$ and the expectation values of the operator
$\calo_\phi(t)$ of the prescribed boundary QFT initial state.
Specifically, extending the asymptotic expansion \eqref{bcdata} for $\{\phi, A\}$,
\begin{equation}
\begin{split}
&\phi=\frac{\Lambda}{r}+\frac{f_2(t)}{r^2}+\calo\left(\frac{1}{r^3}\right)\,,\\
&A=\frac{r^2}{2}+\left(\l-\frac{\dot a }{a }\right)r
+\frac{\l^2}{2}-\frac {\Lambda^2}{8}-\frac{\dot a}{a}\ \l-\dot\l+\frac{1}{r}\left(\mu(t)-\frac \Lambda4 f_2(t)-\frac {\Lambda^2}{4}\l \right)
+\calo\left(\frac{1}{r^2}\right)\,,
\end{split}
\eqlabel{extension}
\end{equation}
the observables of interest can be computed following the
holographic renormalization of the model:
\begin{equation}\begin{split}
2\k^2\ \cale(t)=& -4\mu +\frac{\dot a}{a}\ \Lambda^2+\left(\dd_1\ \Lambda^3+2\dd_2\ \Lambda \frac{(\dot a)^2}{a^2}\right)\,,
\end{split}
\eqlabel{vev1}
\end{equation}
\begin{equation}\begin{split}
2\k^2\ P(t)=& -2\mu +\frac 12 \Lambda ( f_2+\l \Lambda)+\left(-\dd_1\ \Lambda^3-2\dd_2\ \Lambda \frac{\ddot a}{a}\right)\,,
\end{split}
\eqlabel{vev2}
\end{equation}
\begin{equation}
\begin{split}
2\k^2\ \calo_\phi(t)=&-f_2-\l\Lambda +\frac{\dot a}{a}\ \Lambda+\left(3 \dd_1\ \Lambda^2+\dd_2\ \left( 4 \frac{\ddot a}{a}+ 2\frac{(\dot a)^2}{a^2}\right) \right)\,,
\end{split}
\eqlabel{vev3}
\end{equation}
where the terms in brackets, depending on arbitrary constants $\{\dd_1,\dd_2\}$, encode the renormalization scheme 
ambiguities. Independent of the renormalization scheme, 
these expectation values 
satisfy the expected conformal Ward identity
\begin{equation}
\begin{split}
&-\cale+2P=-\Lambda \calo_\phi\,.
\end{split}
\eqlabel{ward}
\end{equation}
Furthermore, the conservation of the stress-energy tensor 
\begin{equation}\eqlabel{cons}
\frac{d\cale}{dt}+2\frac{\dot a}{a} (\cale+P)=0\,,
\end{equation}
is a consequence of the momentum constraint \eqref{mom}:   
\begin{equation}
\begin{split}
&0=\dot \mu+  \frac{\dot a }{a}\left(3\mu-\frac 14 \Lambda f_2\right)-\frac{\Lambda^2}{4}
\biggl(\frac{\dot a}{a}\left(\l+\frac{\dot a}{a}\right)+\frac{\ddot a}{a}\biggr) \,. 
\end{split}
\eqlabel{ward2}
\end{equation}
From now on we choose a scheme with $\dd_i=0$.

One of the advantages of the holographic formulation of a QFT dynamics is the natural definition 
of its far-from-equilibrium entropy density. A gravitational geometry \eqref{EFmetric} 
has an apparent horizon located at $r=r_{AH}$, where \cite{Chesler:2013lia}
\begin{equation}
d_+\Sigma\bigg|_{r=r_{AH}}=0\,.
\eqlabel{defhorloc}
\end{equation} 
Following \cite{Booth:2005qc,Figueras:2009iu} we associate the non-equilibrium  entropy density $s$
of the boundary QFT  with the Bekenstein-Hawking entropy density of the apparent horizon  
\begin{equation}
a^2 s =\frac {2\pi}{\k^2}\ {\Sigma^2}\bigg|_{r=r_{AH}}\,.
\eqlabel{as}
\end{equation}
Using the holographic background equations of motion \eqref{ev1}-\eqref{mom} 
we find 
\begin{equation}
\frac{d(a^2 s)}{dt}=\frac{2\pi}{\k^2}\ (\Sigma^2)'\ \frac{
 (d_+\phi)^2}{\phi^2+6}\bigg|_{r=r_{AH}}\,.
\eqlabel{dasdt}
\end{equation}
Following \cite{Buchel:2017pto} it is easy to  prove that the entropy production rate as defined by \eqref{dasdt}
is non-negative, \ie 
\begin{equation}
\frac{d(a^2 s)}{dt}\ge 0\,,
\eqlabel{dasdt2}
\end{equation}
in holographic dynamics governed by \eqref{ev1}-\eqref{mom}.

The holographic evolution as explained above is implemented in section \ref{nl}, adopting numerical
codes developed in \cite{Bosch:2017ccw,Buchel:2017map}.

\subsection{Bunch-Davies vacua of holographic toy $QFT_3$}\label{bdv}

Following \cite{Buchel:2017pto}, the equations for the late-time attractor of the evolution (a Bunch-Davies vacuum  \cite{Buchel:2017pto})
can be obtained from \eqref{ev1}-\eqref{mom} taking $t\to \infty $ limit with identification
\begin{equation}
\lim_{t\to\infty}\{\phi,A\}(t,r)=\{\phi,A\}_v\,,\qquad \lim_{t\to\infty} \frac{\Sigma(t,r)}{a(t)}=\sigma_v(r)\,.
\eqlabel{ds3vac}
\end{equation}
Introducing a new radial coordinate 
\begin{equation}
x\equiv \frac Hr\,,
\eqlabel{defx}
\end{equation}
and denoting 
\begin{equation}
\phi_v=p(x)\,,\qquad A_v=\frac{H^2}{2x^2}\ g(x)\,,\qquad \sigma_v=\frac {H}{x}\ f(x)\,,
\eqlabel{redefstat}
\end{equation}
we find 
\begin{equation}
\begin{split}
&0=f''+\frac 14 (p')^2\ f\,,\\
&0=p''+\biggl(
(f p^2+12 f' x^2-12 x f+6 f) f x^2
\biggr)^{-1} \biggl(
2 f^2 x^4 (p')^3-f^2 p (p')^2 x^2+(24 x^4 (f')^2\\
&+4 f x^2 (p^2-6 x+6) f'-2 f^2 x (p^2+6)) p'+12 p ((f')^2 x^2-2 f x f'+f^2)
\biggr)\,,
\end{split}
\eqlabel{fpeoms}
\end{equation}
along with an algebraic expression for $g$:
\begin{equation}
g=-\frac{2 f (f p^2+12 f' x^2-12 x f+6 f)}{f^2 (p')^2 x^2-12 (f')^2 x^2+24x f f' -12 f^2}\,.
\eqlabel{geom}
\end{equation}
Vacuum solution has to satisfy the boundary conditions \eqref{bcdata}, and remain nonsingular 
for $x\in (0,x_{AH}]$, where the location of the apparent horizon $x_{AH}$ is determined from 
\cite{Buchel:2017pto}
\begin{equation}
d_+\Sigma(t, x_{AH})=0\qquad \Longleftrightarrow\qquad \biggl(f(x)\ (2 x +g(x)) -g(x) f'(x)\biggr)\bigg|_{x=x_{AH}}=0\,.
\eqlabel{xah}
\end{equation}
 Without loss of generality 
we fix the diffeomorphism parameter $\l$ so that 
\begin{equation}
A_v(x)\bigg|_{x=\frac 13}=0\,.
\eqlabel{lstat}
\end{equation}
We will always have $x_{AH}>\frac 13$. 

It is straightforward to construct an analytic solution to \eqref{fpeoms} as a series expansion in 
conformal symmetry breaking parameter
\begin{equation}
p_1\equiv \frac{\Lambda}{H}\,,
\eqlabel{defp1}
\end{equation} 
\begin{equation}
\begin{split}
&p=p_1\ \frac{x}{1-x}-\frac{x^2 (2 x-1)}{9(x-1)^3}\ p_1^3-
 \frac{x^2 (875 x^3-647 x^2+9 x+51)}{12960(x-1)^5}\ p_1^5+\calo\left(p_1^7\right)\,,
\end{split}
\eqlabel{solvep}
\end{equation}
\begin{equation}
\begin{split}
&f=1-x+ \frac{x(4x-1)}{24(x-1)}\ p_1^2+\frac{ x (4 x-1) (23 x^2-5 x-5)}{3456(x-1)^3}\ p_1^4\\
&+ \frac{x (49618 x^5-46133 x^4+9055 x^3-2745 x^2+3225 x-645)}{6220800(x-1)^5}\ p_1^6+\calo\left(p_1^8\right)\,,
\end{split}
\eqlabel{solvef}
\end{equation}
which determines following \eqref{geom}
\begin{equation}
\begin{split}
&g=(1-3 x) \biggl(1-x
+\frac{x (3 x-1)}{12(x-1)}\ p_1^2+ \frac{(3 x-1) (19 x^2-2 x-5) x}{1728(x-1)^3}\ p_1^4\\
&+\frac{x (3 x-1) (1937 x^4-1196 x^3+54 x^2-204 x+129)}{622080(x-1)^5}\ p_1^6
+\calo\left(p_1^8\right)\biggr)\,.
\end{split}
\eqlabel{solveg}
\end{equation}
From \eqref{xah}, the apparent horizon is located at 
\begin{equation}
\begin{split}
x_{AH}=&1-\frac16  6^{2/3}\ p_1^{2/3}+\frac{1}{12} 6^{1/3}\ p_1^{4/3}+\frac19\ p_1^2-\frac{20401}{622080} 6^{2/3}\ p_1^{8/3}\\
&-\frac{685273}{12441600} 6^{1/3}\ p_1^{10/3}+\frac{40841057}{99532800}\ p_1^4+\calo\left(p_1^{14/3}\right)\,.
\end{split}
\eqlabel{xahpert}
\end{equation}

\begin{figure}[t]
\begin{center}
\psfrag{e}{{$\frac{384}{c}\ \frac{\cale_v}{H^3} $}}
\psfrag{p}{{$\frac{\Lambda}{H}$}}
  \includegraphics[width=4.0in]{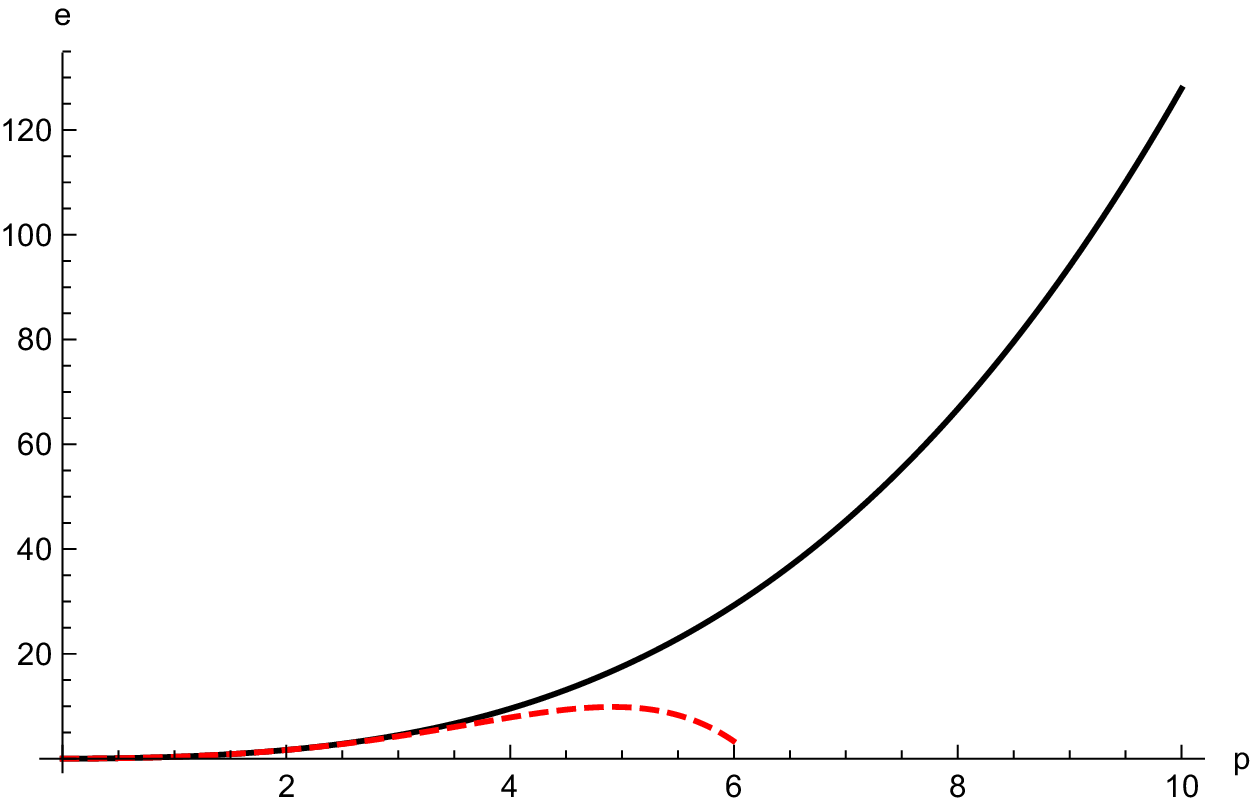}
\end{center}
 \caption{Vacuum energy density $\cale_v$ of the holographic toy model as a function of the conformal symmetry 
breaking deformation $p_1=\frac{\Lambda}{H}$ (solid black line). Dashed red line indicates perturbative prediction, 
see \eqref{enp}.    $c$ is the UV central charge of the model  (see \eqref{c}).  
}\label{figure1}
\end{figure}

For generic $p_1$ we have to resort numerics. Details of the numerical implementation are explained in 
\cite{Buchel:2017pto}. Fig.~\ref{figure1} presents  the vacuum energy $\cale_v$ 
as a function of $p_1$ in renormalization scheme $\dd_i=0$.  Using perturbative solution 
\eqref{solvep}, \eqref{solvef} we find
\begin{equation}
2\k^2\ \frac{\cale_v}{H^3}=\frac13\ p_1^2+\frac{5}{216}\ p_1^4-\frac{43}{51840}\ p_1^6+\calo\left(p_1^8\right)\,.
\eqlabel{enp}
\end{equation}
Note that in vacuum 
$P_v=-\cale_v$, thus  following \eqref{ward},
\begin{equation}
\calo_{\phi,v}=3 \frac {\cale_v}{\Lambda} \,.
\eqlabel{ephiv}
\end{equation}

\begin{figure}[t]
\begin{center}
\psfrag{n}{{$\frac{96}{\pi c}\ \frac{s_{ent}}{H^2} $}}
\psfrag{p}{{$\frac{\Lambda}{H}$}}
  \includegraphics[width=4.0in]{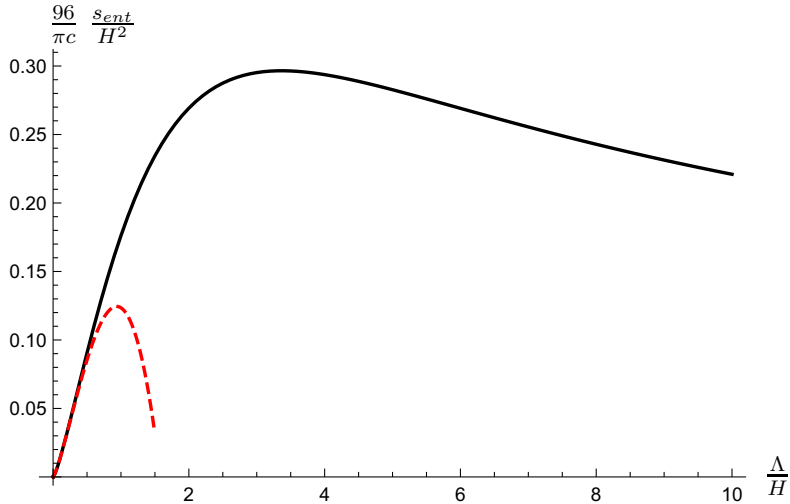}
\end{center}
 \caption{Vacuum entropy density $s_{ent}$ of the holographic toy model (see \eqref{defsent}) as a function of the conformal symmetry 
breaking deformation $p_1=\frac{\Lambda}{H}$. Dashed red line indicates perturbative prediction, 
see \eqref{entp}.  $c$ is the UV central charge of the model  (see \eqref{c}).  
}\label{figure2}
\end{figure}

In  \cite{Buchel:2017qwd} it was argued that the vacuum of a massive QFT in de Sitter has a constant "entanglement'' 
entropy density  $s_{ent}$, related to the comoving entropy production rate $\calr$ at late times. 
Specifically,  parameterizing the comoving entropy production from \eqref{dasdt} as 
\begin{equation}
\lim_{t\to\infty} \frac{1}{H^2 a^2} \frac{d}{dt}\left(a^2 s\right)\equiv 2 H\times \calr\,,
\eqlabel{defrates}
\end{equation}
the vacuum entropy density $s_{ent}$ is
\begin{equation}
s_{ent}\equiv \lim_{t\to \infty} s = H^2\ \calr\,.
\eqlabel{defsent}
\end{equation} 
Fig.~\ref{figure2} presents the vacuum entropy density
as a function of $p_1$ --- this result is renormalization scheme independent. Perturbatively,
\begin{equation}
\begin{split}
\frac{\k^2}{2\pi}\ \frac{s_{ent}}{H^2}=&\frac16 6^{1/3}\ p_1^{4/3}
-\frac{1}{12}\ p_1^2-\frac{5}{216}6^{2/3}\ p_1^{8/3}-\frac{3359}{311040}6^{1/3}\ p_1^{10/3}+\calo\left(p_1^4\right)\,.
\end{split}
\eqlabel{entp}
\end{equation}
Following   \cite{Buchel:2017qwd}, the surface gravity of the apparent horizon equals $(-H)$.

\subsection{Spectrum of vacuum linearized fluctuations}\label{ring}

For static horizons in holography, quasinormal modes of black holes/black branes represent the physical 
linearized fluctuations in the dual boundary field theory plasma at equilibrium. In Fefferman-Graham 
coordinate of the asymptotically AdS bulk geometry the spectrum of QNMs is determined solving Sturm-Liouville
problem for the linearized fluctuations with  Dirichlet conditions at the asymptotic boundary for the non-normalizable 
modes of the fluctuating fields, and 
incoming boundary condition at the horizon \cite{Kovtun:2005ev}. In case of infalling Eddington-Finkelstein 
coordinates (as in \eqref{EFmetric}), the horizon boundary condition is replaced with the regularity at the 
trapped surface (the apparent horizon). We stress again that it is the boundary conditions at the horizon and the 
asymptotic boundary that determine the spectrum of fluctuations. 

In analogy to QNMs, we consider linearized fluctuations of the system  \eqref{ev1}-\eqref{mom} about the 
late-time attractor solution \eqref{ds3vac}. To this end, we define the fluctuations with the harmonic time 
dependence of frequency $\w$ as follows:
\begin{equation}
\begin{split}
&\phi(t,x)=p(x)+\dd\ H_1(x)\ e^{-i \w t}\,,\qquad \frac{\Sigma(t,x)}{a(t)}=\frac{H}{x}\left(f(x)+\dd\ H_2(x)\ e^{-i \w t}\right)\,, \\
&A(t,x)=\frac{H^2}{2x^2}\left(g(x)+\dd\ H_3(x)\ e^{-i \w t}\right)\,,
\end{split}
\eqlabel{fleoms} 
\end{equation}
where $\dd$ is the amplitude of the fluctuations. Substituting \eqref{fleoms} into  \eqref{ev1}-\eqref{mom} and collecting $\calo(\dd)$ terms 
we obtain\footnote{Explicit form of \eqref{fleom1} is provided as a separate file with the arXiv.org submission of this paper.} 
a consistent set of coupled radial equations of motion for $H_i$:
\begin{equation}
\begin{split}
&0=H_1''+ \calc_{1,1}\ H_1'  + \calc_{1,2}\ H_1 + \calc_{1,3}\ H_2     + \calc_{1,4}\ H_3\,,\\
&0=H_2'+ \calc_{2,1}\ H_1'  + \calc_{2,2}\ H_1 + \calc_{2,3}\ H_2     + \calc_{2,4}\ H_3\,,\\    
&0=H_3'+ \calc_{3,1}\ H_1'  + \calc_{3,2}\ H_1 + \calc_{3,3}\ H_2     + \calc_{3,4}\ H_3\,,\\    
\end{split}
\eqlabel{fleom1}
\end{equation} 
where the connection coefficients
\begin{equation}
\calc_{i,j} = \calc_{i,j}\biggl[f'(x),p'(x);\ f(x),p(x);\ x;\ \hw\biggr]
\eqlabel{connection}
\end{equation}
are functionals of vacuum functions $\{f,p\}$ (see \eqref{fpeoms}) and the reduced 
frequency 
\begin{equation}
\hw\equiv \frac\w H\,.
\eqlabel{defhw}
\end{equation}
As in case of the QNMs, we insist that the linearized fluctuations $H_i$ do not change boundary QFT data, \ie we require 
\begin{equation}
H_1=x^2+\calo(x^3)\,,\qquad H_{2}=\calo(x)\,,\qquad H_3=\calo(x)\,,
\eqlabel{bchi}
\end{equation}
as $x\to 0$ (the asymptotic AdS boundary). The $\calo(x^2)$ term in the $H_1$ asymptotic is 
simply the definition of the amplitude of the linearized fluctuations. 
Recall \cite{Buchel:2017pto} that the vacuum equations of motion have a coordinate 
singularity\footnote{There is no coordinate singularity in the radial coordinate in the characteristic formulation of the 
dynamical evolution implemented in section \ref{nl}.} when 
$A_v(x=x_{singularity})=0$. In our case  $x_{singularity}=\frac 13$, see \eqref{lstat}. This coordinate singularity occurs 
{\it always} before the apparent horizon: $x_{AH}> x_{singularity}$. Turns out that the connection coefficients 
$\calc_{i,j}$ are singular at $x_{singualarity}$, and requiring that this is just a coordinate singularity and the fluctuating fields 
$H_i$ are smooth across this point and extend all the way to the apparent horizon $x_{AH}$, provides the second boundary condition 
on the spectrum of fluctuations. 

To recap: the spectrum of linearized fluctuations about Bunch-Davies vacuum is determined from:
\begin{itemize}
\item Dirichlet conditions at the AdS boundary on the non-normalizable modes of the dual gravitational bulk fluctuating fields;
\item regularity condition for bulk fluctuating fields at the location $A_v=0$.    
\end{itemize}

It is instructive to solve \eqref{fleom1} perturbatively in the conformal deformation parameter $p_1$, using 
perturbative expansion for the BD vacuum \eqref{solvep}-\eqref{solvef}. Introducing 
\begin{equation}
H_{i}(x)= \sum_{k=0}^\infty p_1^k\ H_{i,k}(x)\,,\qquad \hw=\sum_{k=0}^\infty\ p_1^k\ \hw_{k}\,,
\eqlabel{perhiw}
\end{equation}
to leading order $k=0$ we find:
\begin{equation}
\begin{split}
&0=H_{1,0}''-\frac{2 i (3 i x^2-\hw_0 x-i)}{x (x-1) (3 x-1)} H_{1,0}'
-\frac{2 i(i x-\hw_0 x-i) }{(3 x-1) (x-1)^2 x^2} H_{1,0} \,,\\
&0=H_{2,0}'+\frac{i (2 i x-i+\hw_0)}{(x-1) (2 x-1)} H_{2,0}-\frac{1}{2x (x-1) (2 x-1)} H_{3,0}\,,\\
&0=H_{3,0}'-\frac{2 \hw_0 x (\hw_0+i)}{(x-1) (2 x-1)} H_{2,0}+\frac{i  (4 i x^2-5 i x-\hw_0 x+i)}{x (x-1) (2 x-1)} H_{3,0}\,.
\end{split}
\eqlabel{k0}
\end{equation}
Note that to leading order in $p_1$ equations for $H_{1}$ and $\{H_2,H_3\}$ decouple. 
The general solution of the first equation in \eqref{k0}, subject to \eqref{bchi}, is 
\begin{equation}
H_{1,0}=\begin{cases}
&-\frac{x}{2(1-x)(1-i \hw_0)}\ \left(1-\left(\frac{1-3 x}{1-x}\right)^{-1+i\hw_0}\right)\,,\qquad \hw_0\ne -i\\
&-\frac{x}{2(1-x)}\ \ln\frac{1-3x}{1-x}\,,\qquad \hw_0= -i\,.
\end{cases}
\eqlabel{h10sol}
\end{equation}
Requiring that $H_{1,0}$ is analytic at $x=x_{singularity}=\frac 13$ produces the spectrum of fluctuations 
to leading order in $p_1$:
\begin{equation}
\hw\equiv \hw^{(n)}= -i n+\calo(p_1)\,,\qquad n=2,3,\cdots
\eqlabel{splead}
\end{equation} 
Note that in a conformal limit $p_1\to 0$ the mode \eqref{h10sol} disappears from the spectrum --- all $(n)$-modes
are singular at $x=x_{AH}=1+\calo(p_1^{2/3})$. We interpreted this fact as a statement that 
{\it the Bunch-Davies vacuum of a CFT does not ring}.   
It is straightforward to check that the remaining two equations in \eqref{k0} do not lead to new spectral 
branches\footnote{This is also confirmed comparing with the relaxation to BD vacuum in the full nonlinear 
dynamics as explained in section \ref{nl}. }. 

The leading order solution \eqref{splead} can be extended to higher orders in $\calo(p_1)$. 
For example, for $n=2$ mode we find:
\begin{equation}
\begin{split}
&\hw^{(2)}=-i\left(2 +\frac{1}{12}\ p_1^2-\frac{1}{54}\ p_1^4
+\frac{1591}{622080}\ p_1^6+\calo\left(p_1^8\right)\right)\,,\\
&H_1^{(2)}=\frac{x^2}{(1-x)^2}+\frac{x^3 (19 x-6)}{36(x-1)^4}\ p_1^2+
\frac{x^3 (6768 x^3-3125 x^2-950 x+535)}{25920(x-1)^6}\ p_1^4 
\\
&+\frac{x^3 (17864583 x^5-14152740 x^4+1089102 x^3-157220 x^2+1292795 x-409080)}{130636800(x-1)^8}\ p_1^6
\\&+\calo\left(p_1^8\right)\,,\\
&H_2^{(2)}=- \frac{x (13 x^2-2 x+1)}{72(x-1)^2}\ p_1
-\frac{x (2867 x^4-1644 x^3+646 x^2-444 x+111)}{34560 (x-1)^4}\  p_1^3 \\
&-\frac{x (17822851 x^6-8946582 x^5-7415409 x^4+8763460 x^3-5161395 x^2+2036586 x-339431)}
{522547200(x-1)^6}\ p_1^5\\
&+\calo\left(p_1^7\right)\,,\\
&H_3^{(2)}=(3x-1)\biggl(
\frac{(4 x^2+x+1) x}{36(x-1)^2}\ p_1+\frac{(1036 x^4+95 x^3-447 x^2-51 x+111) x}{17280(x-1)^4}\ p_1^3
\\
&+\frac{
(8634226 x^6-6320265 x^5+1435341 x^4-731078 x^3-568056 x^2+1083183 x-339431) x}{261273600(x-1)^6}\ p_1^5\\
&+\calo\left(p_1^7\right)\,,
\biggr)
\end{split}
\eqlabel{mode2}
\end{equation}
where we fixed the  diffeomorphism parameter $\l(t)$ to all orders in $p_1$ requiring that $A(t,x=\frac 13)=0$ .
Note that $n=2$ mode is purely dissipative. In fact, we find that all modes except for $n=3$ are purely dissipative.
For example, 
\begin{equation}
\begin{split}
&\hw^{(4)}=-i\left(4-\frac{1}{216}\ p_1^4+\frac{337}{777600}\ p_1^6+\calo\left(p_1^8\right)\right)\,,\\
&\hw^{(5)}=-i\left(5-\frac{5}{4096}\ p_1^4-\frac{589}{15925248} p_1^6+\calo\left(p_1^8\right)\right)\,,\\
\end{split}
\eqlabel{mode45}
\end{equation}
while 
\begin{equation}
\begin{split}
&\hw^{(3)}=-3 i+\frac14  \sqrt{2}\ p_1-\frac{11i}{192}\  p_1^2+\frac{37}{12288} \sqrt{2}\ p_1^3 
-\frac{1855i}{221184}\  p_1^4-\frac{2076503}{1132462080} \sqrt{2}\  p_1^5\\
&+\frac{1240993i}{1061683200}\ p_1^6+\calo\left(p_1^7\right)\,.\\
\end{split}
\eqlabel{mode3}
\end{equation}
What makes the mode $n=3$ special is the fact that the connection coefficients $\calc_{i,j}$ in \eqref{fleom1} have 
a simple pole at $\hw=-3 i$. Unfortunately, we do not understand the physical reason for this.

\begin{figure}[t]
\begin{center}
\psfrag{b}{{$-\Im{[\hw^{(n)}]} $}}
\psfrag{r}{{$\Re{[\hw^{(3)}]} $}}
\psfrag{p}{{$\frac{\Lambda}{H}$}}
  \includegraphics[width=2.8in]{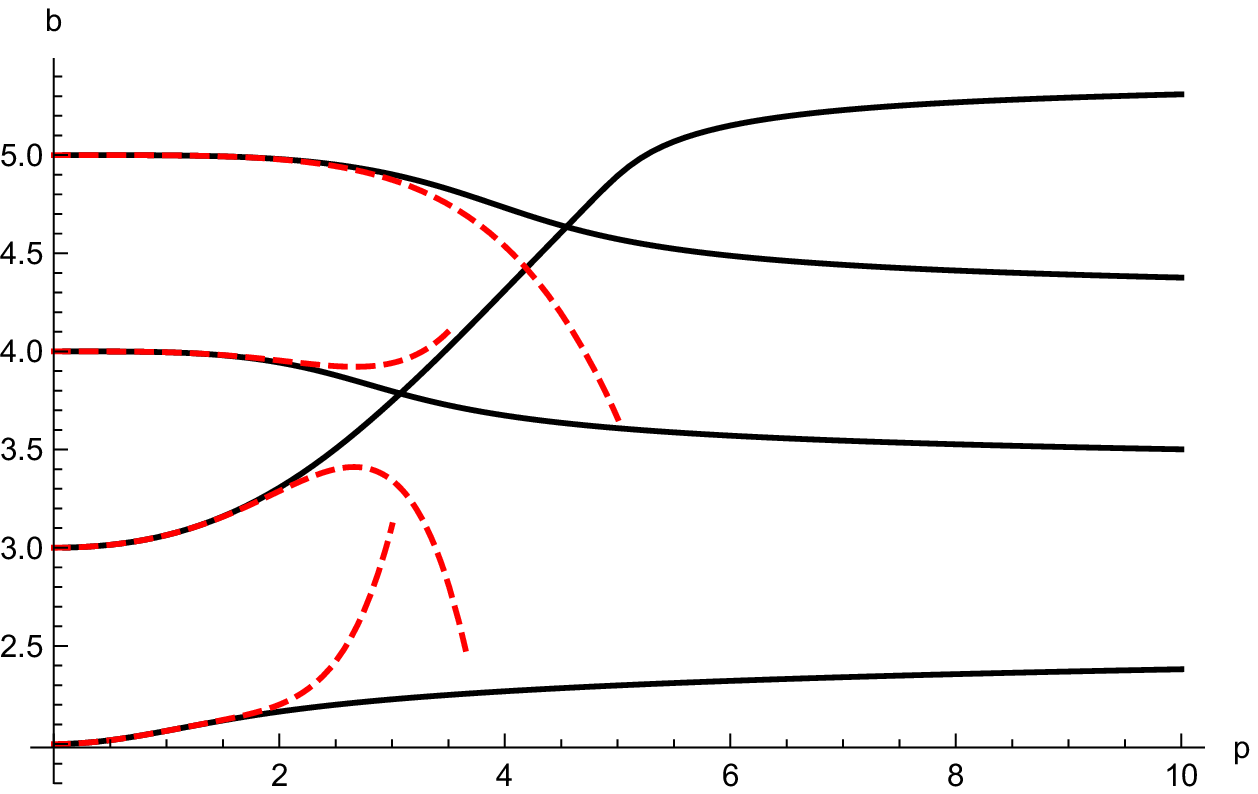}
  \includegraphics[width=2.8in]{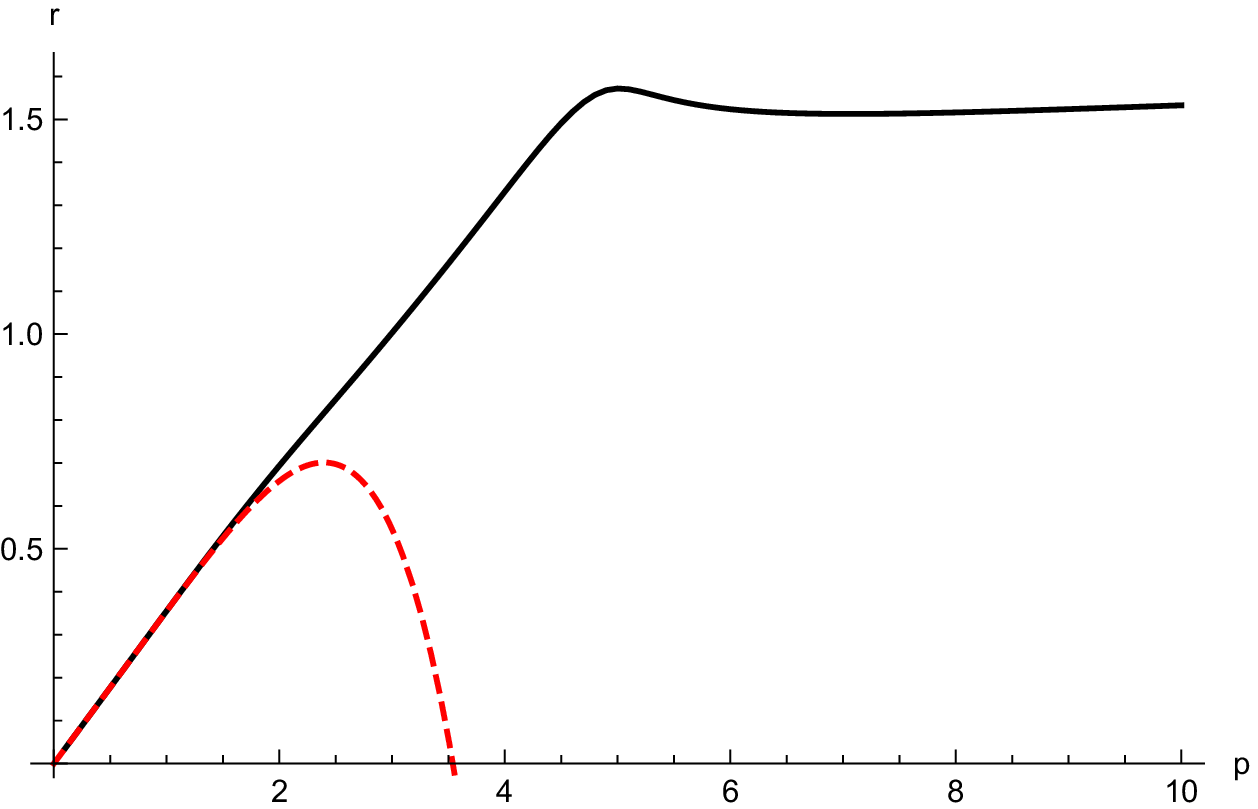}
\end{center}
 \caption{Spectra of BD vacuum fluctuations for several low-lying modes $n=2,3,4,5$.
Red dashed lines   indicate perturbative predictions perturbative 
 in conformal symmetry breaking parameter $p_1=\frac{\Lambda}{H}$. 
}\label{figure3}
\end{figure}

For general $p_1$ the spectrum of fluctuations can be computed numerically. These results are presented in 
fig.~\ref{figure3} for $n=\{2,3,4,5\}$ modes. The dashed red lines indicate perturbative approximation \eqref{mode2},
\eqref{mode45} and \eqref{mode3}. In what follows to refer to the fluctuations in BD vacuum as QNMs.

\begin{figure}[t]
\begin{center}
\psfrag{x}{{$x-x_{singularity}$}}
\psfrag{h}{{$H_{1}^{(2)}$}}
  \includegraphics[width=4.0in]{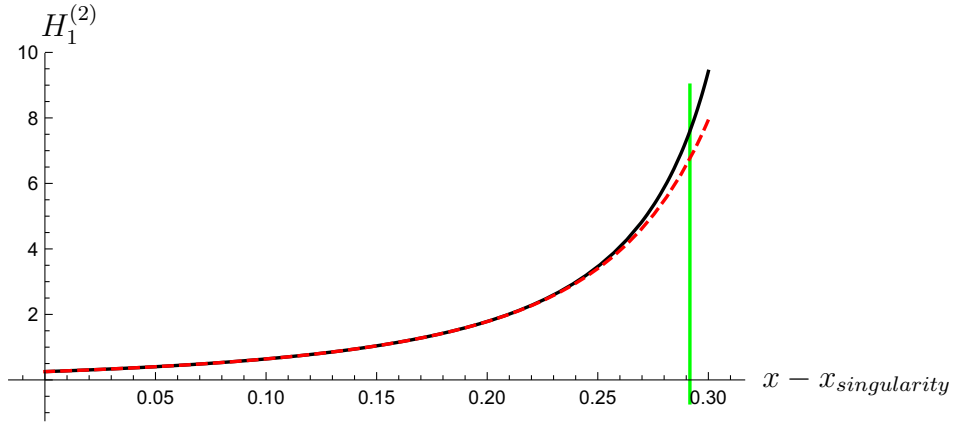}
\end{center}
 \caption{ Fluctuation profile $H_1^{(2)}$ at $p_1=1$ as a function of $x>x_{singularity}$. The vertical 
 green  line indicates the location of the apparent horizon, see \eqref{xah}. The dashed red line is the 
perturbative prediction \eqref{mode2}.  
}\label{figure4}
\end{figure}

Although the spectrum is determined solving \eqref{fleom1} on the radial interval 
$x\in (0,x_{singularity})$, we verified that the solution can indeed be smoothly extended
to the full interval $x\in (0,x_{AH})$. For example, the radial profile  $H_{1}^{(2)}(x)$ at $p_1=1$ 
is presented in fig.~\ref{figure4}.

\subsection{Fully nonlinear dynamics and relaxation to BD vacuum}\label{nl}

In this section we report results of the fully nonlinear evolution of the toy holographic QFT defined by a dual 
gravitational action \eqref{s4}.  Numerical implementation parallel the codes developed in \cite{Bosch:2017ccw,Buchel:2017map},
and will not be discussed here. 

In what follows we focus on the model\footnote{The discussion is generic 
for the parameter set with stable and convergent evolution of the code.} with $p_1=1$. We use the radial coordinate as in \eqref{defx}
and evolve in dimensionless time $\t\equiv H t$. As in \cite{Chesler:2013lia} we adjust the diffeomorphism parameter 
$\l(t)$ so that the apparent horizon is always at $x_{AH}=1$. 
 We set the initial condition for the evolution as 
\begin{equation}
\phi(0,x)=\phi_{initial}(x)=p_1 x+ \cala\ x^2 e^{-x} \,,
\eqlabel{initla}
\end{equation}
where $\cala$ is the amplitude. 
We also need to supply the initial energy density (see \eqref{ward2})
\begin{equation}
\mu(0)\equiv \mu_{initial}\,.
\eqlabel{muinitial}
\end{equation}
We verified that BD vacuum is indeed the attractor of long-time dynamics by choosing different 
initial states for the evolution, \ie different profiles $\phi_{initial}$ and/or $\mu_{initial}$.  

\begin{figure}[t]
\begin{center}
\psfrag{t}{{$\tau$}}
\psfrag{v}{{$\frac{384}{c}\ \frac{\calo_\phi(\t)}{H^2}-\frac{\Lambda}{H}$}}
  \includegraphics[width=4.0in]{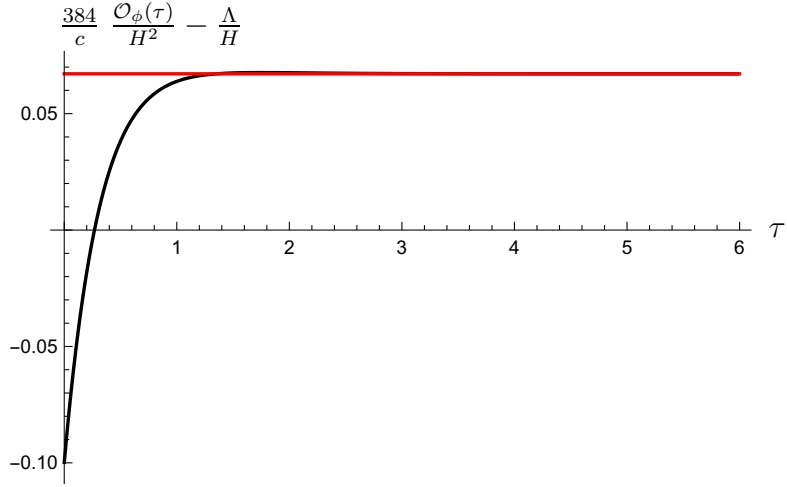}
\end{center}
 \caption{ Fully nonlinear dynamical evolution of the initial boundary QFT state \eqref{initla}
(the solid black curve) as a function of $\t=H t$. The red line represents the late-time asymptotic value of the operator 
$\calo_\phi$ computed in BD vacuum in section \ref{bdv}.
}\label{figure5}
\end{figure}

\begin{figure}[t]
\begin{center}
\psfrag{t}{{$\tau$}}
\psfrag{v}{{$\dd=|1-{\calo_\phi^{QNM}(\t)}/{\calo_\phi(\t)}|$}}
  \includegraphics[width=2.8in]{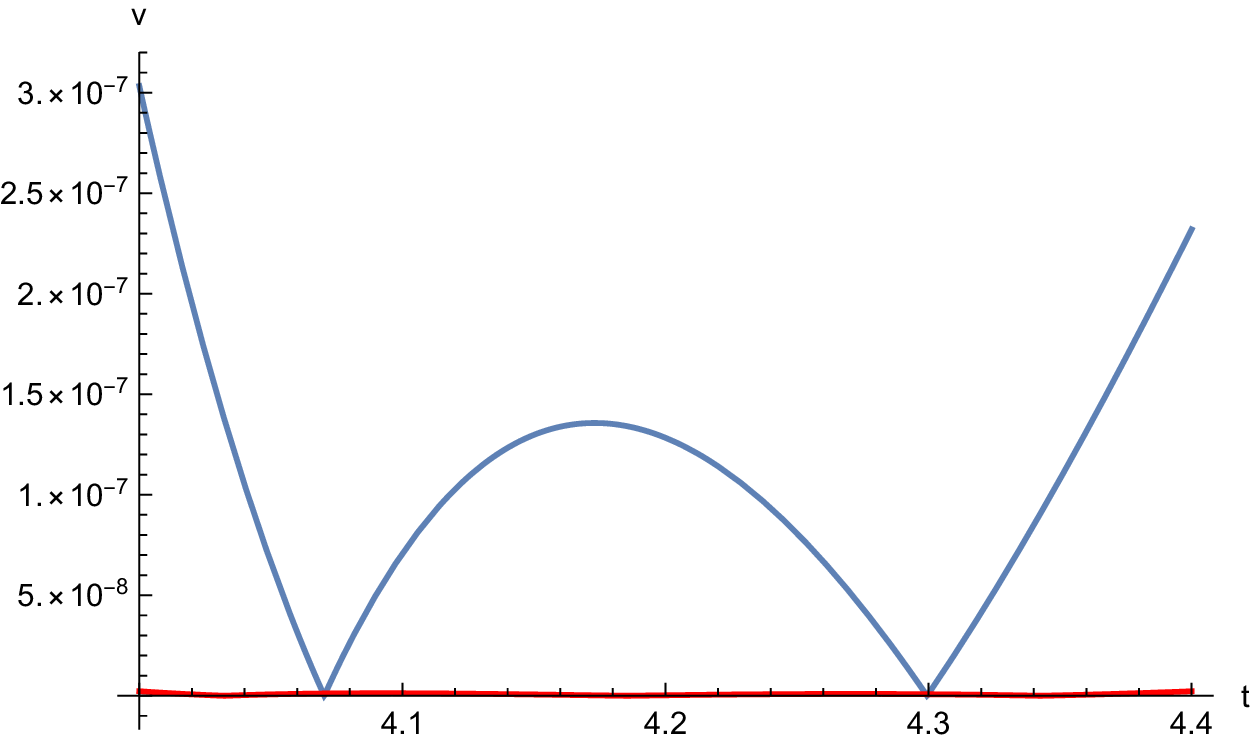}
  \includegraphics[width=2.8in]{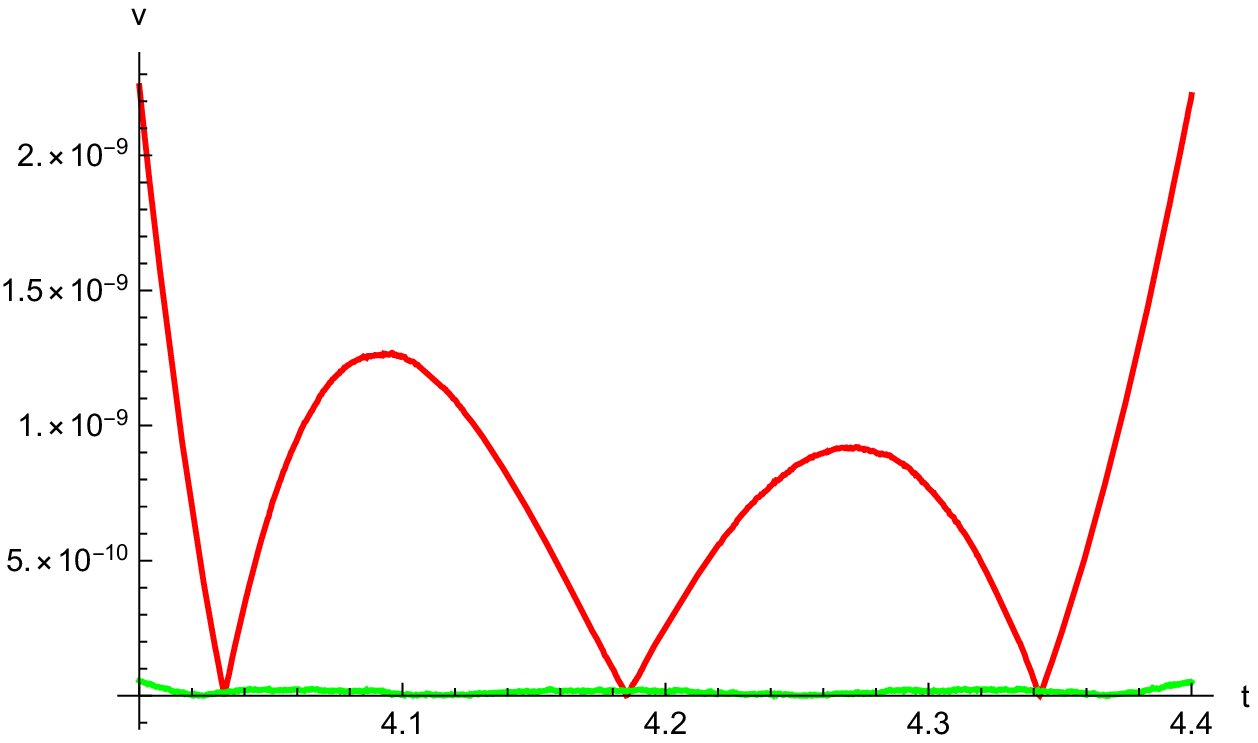}
\end{center}
 \caption{Relaxation to BD vacuum via the QNMs computed in section \ref{ring}. The residuals $\dd$ are computed with the 
best fit $\calo_{\phi}^{QNM}$ to the evolution data using a constant and:  a single $n=2$ mode (blue curve); two modes 
$n=2,3$ (red curve); 3 modes $n=2,3,4$ (green curve).  
}\label{figure6}
\end{figure}

Fig.~\ref{figure5} represent a typical dynamical evolution of the boundary QFT state from the initial condition  \eqref{initla}.
As times $\t=H t\gtrsim 2 $ the state relaxes to BD vacuum. The relaxation process is studied in further details as follows.
\nxt We use the last 1000 data points, corresponding to time interval $\t\in [5.6,6]$ and fit the observed $\calo_\phi(\t)$
with a single QNM ansatz:
\begin{equation}
\calo_{\phi}^{fit} = \a_1+ \a_2 e^{-i \a_3 \t}\,,
\eqlabel{fit1} 
\end{equation}
where $\a_i$ are constant free parameters. $\a_1$ is expected to agree with the BD expectation value and $\a_3$ should 
approximate the frequency of the lowest BD QNM mode, \ie $n=2$. We find that the BD vacuum expectation value is correct with 
a relative error of $\sim 3 \times 10^{-6}$ and the relative error in the frequency, 
\begin{equation}
\bigg|1-\frac{\a_3}{\hw^{(2)}}\bigg|= 2.7\times 10^{-3}\,,
\eqlabel{al3}
\end{equation}
is in excellent agreement with the result in section \ref{ring}.  
\nxt To check on the spectrum of higher QNMs computed in  section \ref{ring} 
we restrict to 1000 data points in the intermediate time-range, 
$\t\in [4,4.4]$. We compute residual $\dd$ defined as 
\begin{equation}
\dd(\t)\equiv \bigg|1-\frac{\calo_\phi^{QNM}(\t)}{\calo_\phi(\t)}\bigg|\,,
\eqlabel{deldd}
\end{equation}
where $\calo_\phi^{QNM}(\t)$ is the best QNM approximation to the data in the time subinterval with constant free 
fit parameters $\a_i$ and frequencies $\hw^{(n)}$ computed in section \ref{ring}:
\begin{equation}
\begin{split}
&n=2:\qquad \calo_\phi^{QNM}=\a_1 + \a_2 e^{-i \hw^{(2)} \t}\,, \\
&n=2,3:\qquad \calo_\phi^{QNM}=\a_1 + \a_2 e^{-i \hw^{(2)} \t}+\a_3 e^{\Im [\hw^{(3)}]\t}\cos[\Re[\hw^{(3)}]\t +\a_4]\,,  \\
&n=2,3,4:\qquad \calo_\phi^{QNM}=\a_1 + \a_2 e^{-i \hw^{(2)} \t}+\a_3 e^{\Im [\hw^{(3)}]\t}\cos[\Re[\hw^{(3)}]\t +\a_4]
   + \a_5 e^{-i \hw^{(4)} \t}\,.
\end{split}
\eqlabel{oqnm}
\end{equation}
The residual $\dd$ is presented in fig.~\ref{figure6}. The quality of approximation suggests that the QNMs computed 
in section \ref{ring} are {\it all } the modes defining the relaxation of the theory to its BD vacuum.

\section{Conclusion}\label{end}  
A surprising fact discovered in \cite{Buchel:2017pto,Buchel:2017qwd} is that a vacuum of a massive QFT in de Sitter space-time has a constant 
entropy density $s_{ent}$. We stress that it is important that both the Hubble constant is nonzero, and that the theory is non-conformal.
For example, in a simple 2+1 dimensional holographic toy model discussed here 
\begin{equation}
s_{ent} \sim c\ \Lambda^{4/3} H^{2/3}\,,\qquad \frac{\Lambda}{H}\ll 1\,,
\eqlabel{ssmall}
\end{equation}
where $c$ is a UV central charge of the model and $\Lambda$ is a mass scale of the theory. 

Thermal equilibrium states have entropy. (Non)-hydrodynamic modes in equilibrium plasma owe their existence to
this entropy --- no entropy, nothing to excite. By analogy, the nonvanishing vacuum entropy of a  massive QFT in de Sitter 
suggests that there should be analogous QNM-like excitations about its Bunch-Davies vacuum. 
In this paper we showed that this is indeed the case.

\section*{Acknowledgments}
Research at Perimeter
Institute is supported by the Government of Canada through Industry
Canada and by the Province of Ontario through the Ministry of
Research \& Innovation. This work was further supported by
NSERC through the Discovery Grants program.



\begin{thebibliography}{99}


\bibitem{mbl} R.~Nandkishore and D.~A.~Huse, 
``Many-body localization and thermalization in quantum statistical mechanics,''
Annu. Rev. Condens. Matter Phys. {\bf 6.1}, 15 (2015).  

\bibitem{Balasubramanian:2014cja} 
  V.~Balasubramanian, A.~Buchel, S.~R.~Green, L.~Lehner and S.~L.~Liebling,
  ``Holographic Thermalization, Stability of Anti–de Sitter Space, and the Fermi-Pasta-Ulam Paradox,''
  Phys.\ Rev.\ Lett.\  {\bf 113}, no. 7, 071601 (2014)
  doi:10.1103/PhysRevLett.113.071601
  [arXiv:1403.6471 [hep-th]].


\bibitem{Baier:2007ix} 
  R.~Baier, P.~Romatschke, D.~T.~Son, A.~O.~Starinets and M.~A.~Stephanov,
  ``Relativistic viscous hydrodynamics, conformal invariance, and holography,''
  JHEP {\bf 0804}, 100 (2008)
  doi:10.1088/1126-6708/2008/04/100
  [arXiv:0712.2451 [hep-th]].


\bibitem{Bhattacharyya:2008jc} 
  S.~Bhattacharyya, V.~E.~Hubeny, S.~Minwalla and M.~Rangamani,
  ``Nonlinear Fluid Dynamics from Gravity,''
  JHEP {\bf 0802}, 045 (2008)
  doi:10.1088/1126-6708/2008/02/045
  [arXiv:0712.2456 [hep-th]].

\bibitem{Heller:2013fn} 
  M.~P.~Heller, R.~A.~Janik and P.~Witaszczyk,
  ``Hydrodynamic Gradient Expansion in Gauge Theory Plasmas,''
  Phys.\ Rev.\ Lett.\  {\bf 110}, no. 21, 211602 (2013)
  doi:10.1103/PhysRevLett.110.211602
  [arXiv:1302.0697 [hep-th]].


\bibitem{Buchel:2016cbj} 
  A.~Buchel, M.~P.~Heller and J.~Noronha,
  ``Entropy Production, Hydrodynamics, and Resurgence in the Primordial Quark-Gluon Plasma from Holography,''
  Phys.\ Rev.\ D {\bf 94}, no. 10, 106011 (2016)
  doi:10.1103/PhysRevD.94.106011
  [arXiv:1603.05344 [hep-th]].




\bibitem 
 {m1} J.~M.~Maldacena,
  ``The large N limit of superconformal field theories and supergravity,''
  Adv.\ Theor.\ Math.\ Phys.\  {\bf 2}, 231 (1998)
  [Int.\ J.\ Theor.\ Phys.\  {\bf 38}, 1113 (1999)]
  [arXiv:hep-th/9711200].


\bibitem{Aharony:1999ti} 
  O.~Aharony, S.~S.~Gubser, J.~M.~Maldacena, H.~Ooguri and Y.~Oz,
  ``Large N field theories, string theory and gravity,''
  Phys.\ Rept.\  {\bf 323}, 183 (2000)
  [hep-th/9905111].

\bibitem{Witten:1998zw} 
  E.~Witten,
  ``Anti-de Sitter space, thermal phase transition, and confinement in gauge theories,''
  Adv.\ Theor.\ Math.\ Phys.\  {\bf 2}, 505 (1998)
  [hep-th/9803131].


\bibitem{Berti:2009kk} 
  E.~Berti, V.~Cardoso and A.~O.~Starinets,
  ``Quasinormal modes of black holes and black branes,''
  Class.\ Quant.\ Grav.\  {\bf 26}, 163001 (2009)
  doi:10.1088/0264-9381/26/16/163001
  [arXiv:0905.2975 [gr-qc]].


\bibitem{Buchel:2015saa} 
  A.~Buchel, M.~P.~Heller and R.~C.~Myers,
  ``Equilibration rates in a strongly coupled nonconformal quark-gluon plasma,''
  Phys.\ Rev.\ Lett.\  {\bf 114}, no. 25, 251601 (2015)
  doi:10.1103/PhysRevLett.114.251601
  [arXiv:1503.07114 [hep-th]].


\bibitem{Fuini:2015hba} 
  J.~F.~Fuini and L.~G.~Yaffe,
  ``Far-from-equilibrium dynamics of a strongly coupled non-Abelian plasma with non-zero charge density or external magnetic field,''
  JHEP {\bf 1507}, 116 (2015)
  doi:10.1007/JHEP07(2015)116
  [arXiv:1503.07148 [hep-th]].


\bibitem{Janik:2015waa} 
  R.~A.~Janik, G.~Plewa, H.~Soltanpanahi and M.~Spalinski,
  ``Linearized nonequilibrium dynamics in nonconformal plasma,''
  Phys.\ Rev.\ D {\bf 91}, no. 12, 126013 (2015)
  doi:10.1103/PhysRevD.91.126013
  [arXiv:1503.07149 [hep-th]].


\bibitem{Buchel:2015ofa} 
  A.~Buchel and A.~Day,
  ``Universal relaxation in quark-gluon plasma at strong coupling,''
  Phys.\ Rev.\ D {\bf 92}, no. 2, 026009 (2015)
  doi:10.1103/PhysRevD.92.026009
  [arXiv:1505.05012 [hep-th]].

\bibitem{Buchel:2002wf} 
  A.~Buchel,
  ``Gauge / gravity correspondence in accelerating universe,''
  Phys.\ Rev.\ D {\bf 65}, 125015 (2002)
  doi:10.1103/PhysRevD.65.125015
  [hep-th/0203041].


\bibitem{Buchel:2002kj} 
  A.~Buchel, P.~Langfelder and J.~Walcher,
  ``On time dependent backgrounds in supergravity and string theory,''
  Phys.\ Rev.\ D {\bf 67}, 024011 (2003)
  doi:10.1103/PhysRevD.67.024011
  [hep-th/0207214].

\bibitem{Buchel:2003qm} 
  A.~Buchel,
  ``Compactifications of the N = 2* flow,''
  Phys.\ Lett.\ B {\bf 570}, 89 (2003)
  doi:10.1016/j.physletb.2003.07.030
  [hep-th/0302107].

\bibitem{Buchel:2004qg} 
  A.~Buchel and A.~Ghodsi,
  ``Braneworld inflation,''
  Phys.\ Rev.\ D {\bf 70}, 126008 (2004)
  doi:10.1103/PhysRevD.70.126008
  [hep-th/0404151].


\bibitem{Buchel:2006em} 
  A.~Buchel,
  ``Inflation on the resolved warped deformed conifold,''
  Phys.\ Rev.\ D {\bf 74}, 046009 (2006)
  doi:10.1103/PhysRevD.74.046009
  [hep-th/0601013].


\bibitem{Buchel:2013dla} 
  A.~Buchel and D.~A.~Galante,
  ``Cascading gauge theory on $dS_4$ and String Theory landscape,''
  Nucl.\ Phys.\ B {\bf 883}, 107 (2014)
  doi:10.1016/j.nuclphysb.2014.03.022
  [arXiv:1310.1372 [hep-th]].



\bibitem{Buchel:2017pto} 
  A.~Buchel and A.~Karapetyan,
  ``de Sitter Vacua of Strongly Interacting QFT,''
  JHEP {\bf 1703}, 114 (2017)
  doi:10.1007/JHEP03(2017)114
  [arXiv:1702.01320 [hep-th]].

\bibitem{Buchel:2017qwd} 
  A.~Buchel,
  ``Verlinde Gravity and AdS/CFT,''
  arXiv:1702.08590 [hep-th].

\bibitem{Chesler:2013lia} 
  P.~M.~Chesler and L.~G.~Yaffe,
  ``Numerical solution of gravitational dynamics in asymptotically anti-de Sitter spacetimes,''
  JHEP {\bf 1407}, 086 (2014)
  doi:10.1007/JHEP07(2014)086
  [arXiv:1309.1439 [hep-th]].

\bibitem{Booth:2005qc} 
  I.~Booth,
  ``Black hole boundaries,''
  Can.\ J.\ Phys.\  {\bf 83}, 1073 (2005)
  doi:10.1139/p05-063
  [gr-qc/0508107].

\bibitem{Figueras:2009iu} 
  P.~Figueras, V.~E.~Hubeny, M.~Rangamani and S.~F.~Ross,
  ``Dynamical black holes and expanding plasmas,''
  JHEP {\bf 0904}, 137 (2009)
  doi:10.1088/1126-6708/2009/04/137
  [arXiv:0902.4696 [hep-th]].

\bibitem{Bosch:2017ccw} 
  P.~Bosch, A.~Buchel and L.~Lehner,
  ``Unstable horizons and singularity development in holography,''
  arXiv:1704.05454 [hep-th].

\bibitem{Buchel:2017map} 
  A.~Buchel,
  ``Singularity development and supersymmetry in holography,''
  arXiv:1705.08560 [hep-th].

\bibitem{Kovtun:2005ev} 
  P.~K.~Kovtun and A.~O.~Starinets,
  ``Quasinormal modes and holography,''
  Phys.\ Rev.\ D {\bf 72}, 086009 (2005)
  doi:10.1103/PhysRevD.72.086009
  [hep-th/0506184].


\end{thebibliography}
\end{document}